\documentclass[onecollarge]{svjour2}
\smartqed  
\usepackage{graphicx}
\newcommand{\be}{\begin{equation}}
\newcommand{\ee}{\end{equation}}
\newcommand{\ba}{\begin{eqnarray}}
\newcommand{\ea}{\end{eqnarray}}
\newcommand{\nn}{\nonumber}
\newcommand{\ci}[1]{\cite{#1}}
\newcommand{\gev}{\,{\rm GeV}}

\newcommand{\tw}{\textwidth}
\newcommand{\req}[1]{(\ref{#1})}

\newcommand{\gsim}{\raisebox{-4pt}{$\,\stackrel{\textstyle
      >}{\sim}\,$}}

\journalname{Archive of Applied Mechanics}
\begin{document}

\title{Hard exclusive pion leptoproduction}

\author{Peter Kroll}


\institute{P. Kroll \at  
           Fachbereich Physik, Universit\"at Wuppertal, 42097 Wuppertal\\
              Tel.: +49-202-4392620\\
              Fax: +49-202-4393811\\
              \email{pkroll@uni-wuppertal.de } }

\date{Received: date / Accepted: date}

\maketitle

\begin{abstract}
In this talk it is reported on an analysis of hard exclusive  
leptoproduction of pions within the handbag approach. It is argued that 
recent measurements of this process performed by HERMES and CLAS clearly 
indicate the occurrence of strong contributions from transversely polarized 
photons. Within the handbag approach such $\gamma^{\,*}_T\to \pi$ transitions 
are described by the transversity GPDs accompanied by twist-3 pion wave 
functions. It is shown that the handbag approach leads to results on cross 
sections and single-spin asymmetries in fair agreement with experiment. 
Predictions for other pseudoscalar meson channels are also briefly discussed.
\end{abstract}

\section{Introduction}
\label{intro}
The handbag approach to hard exclusive leptoproduction of mesons offers a
partonic description of these processes in the generalized Bjorken regime
of large photon virtuality, $Q^2$, and large photon-proton center of mass 
energy, $W$, but small invariant momentum transfer, $-t$. The theoretical 
basis of the handbag approach is provided by the factorization theorems
\ci{rad96,collins96} which say that the process amplitudes for 
longitudinally polarized virtual photons, $\gamma_L^*$, are represented as 
convolutions of hard partonic subprocess amplitudes and soft hadronic 
matrix elements, so-called generalized parton distributions (GPDs), which 
encode the soft physics. The subprocess amplitudes are likewise convolutions 
of perturbatively calculable hard scattering kernels and meson wave 
functions. For pion production in particular the helicity amplitudes for 
$\gamma_L^*\to\pi$ transitions read
\be
{\cal M}_{0+,0+}\,=\, \frac{e_0}{Q}\sqrt{1-\xi^2} \Big[ \langle \widetilde H \rangle
                    - \frac{\xi^2}{1-\xi^2} \langle \tilde E \rangle \Big]\,, 
\qquad {\cal M}_{0-,0+}\,=\, \frac{e_0}{Q} \frac{\sqrt{-t'}}{2m} \xi 
                                                 \langle \tilde E \rangle \,, 
\label{eq:twist2}
\ee
where 
\be
\langle K \rangle \,=\,\int^1_{-1} dx \sum_\lambda 
                             {\cal H}_{0\lambda,0\lambda}(x,\xi,Q^2) K(x,\xi,t)
\ee 
denotes the convolution of a subprocess amplitude, ${\cal H}$, with a GPD 
$K$. The nucleon mass is denoted by $m$. The skewness, $\xi$, is related to 
Bjorken-x by $\xi =x_B/(1-x_B)$ up to corrections of order $1/Q^2$. In 
\req{eq:twist2} the usual abbreviation $t'=t-t_0$ is employed where the minimal 
value of $-t$ corresponding to forward scattering, is $t_0=-4m^2\xi^2/(1-\xi^2)$.
Helicities are labeled by their signs or by zero; they appear in the familiar 
order: pion, outgoing nucleon, photon, in-going nucleon.

Power corrections to the leading-twist result \req{eq:twist2} are theoretically 
not under control. It is therefore not clear at which values of $Q^2$ and $W$ 
the amplitudes \req{eq:twist2} can be applied. The onset of the leading-twist
dominance has to be found out by comparison with experiment. An example of power 
corrections is set by the amplitudes for transversely polarized photons, 
$\gamma_T^*$, which are suppressed by $1/Q$ as compared to the $\gamma_L^*\to\pi$ 
amplitudes. Still, as experiments tell us, the $\gamma_T^*\to\pi$ amplitudes play 
an important role  in hard exclusive pion leptoproduction. The first experimental 
evidence for strong contribution from $\gamma_T^*\to\pi$ transitions came from the 
spin asymmetry, $A_{UT}$, measured with a transversely polarized target by the HERMES 
collaboration for $\pi^+$ production \ci{hermes-aut}. Its $\sin{\phi_s}$ 
modulation~\footnote{
           The angle $\phi_s$ defines the orientation of the target spin vector.}
unveils a particularly striking behavior: It is rather large and does not show any 
indication of a turnover towards zero for $t'\to 0$. In this limit 
$A_{UT}^{\sin \phi_s}$ is in control of an interference term of two helicity non-flip 
amplitudes 
\be
 A_{UT}^{\sin \phi_s}\propto {\rm Im} \Big[{\cal M}^*_{0-,++}{\cal M}_{0+,0+}\Big]\,.
\ee
Both the amplitudes are not forced by angular momentum conservation to vanish in 
the limit $t'\to 0$. Hence, the small $-t'$ behavior of $A_{UT}^{\sin \phi_s}$ entails
a sizeable $\gamma_T^+\to\pi$ amplitude.
  
\begin{figure*}
\centering
\includegraphics[width=0.38\tw]{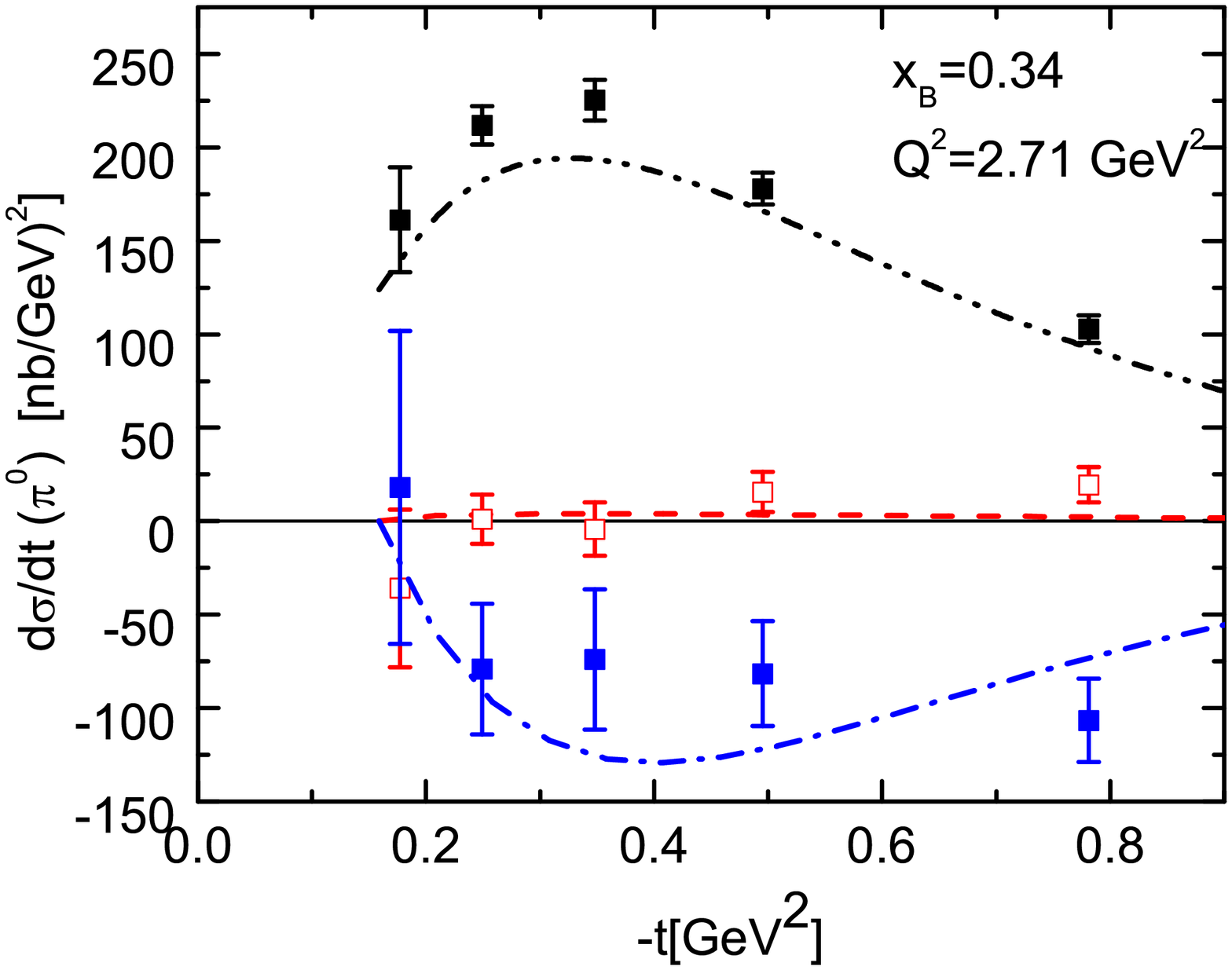}\hspace*{0.05\tw}
\includegraphics[width=0.46\tw]{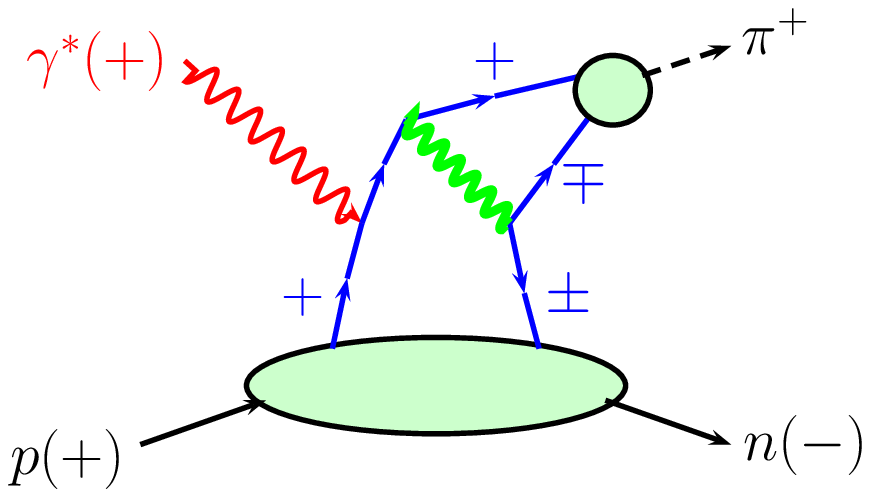}
\caption{\label{fig:1} The unseparated (long.-transv., transv.-transv.) cross 
section. Data \ci{clas-pi0} are shown as solid (open, solid) symbols. The 
theoretical results are taken from \ci{GK6}.}
\caption{\label{fig:2} A typical lowest-order Feynman graph for pion 
leptoproduction. The signs indicate the helicities of the involved particles. }      
\end{figure*}

A second evidence for large contributions from transversely polarized 
photons comes from the CLAS measurement \ci{clas-pi0} of the $\pi^0$ 
electroproduction cross sections. As can be seen from Fig.\ \ref{fig:1} the 
transverse-transverse interference cross section is negative and amounts to about
$50\%$ of the unseparated cross section in absolute value. Neglecting the
double-flip amplitude ${\cal M}_{0-,-+}$ ( $\propto t'$ for $t'\to 0$) and 
introducing the combinations
\be 
{\cal M}^{N,U}_{0+,\mu +}\,=\,\frac12 \Big[{\cal M}_{0+,\mu +} \pm 
                                     {\cal M}_{0+,-\mu +}\Big]\,,
\label{eq:N-U}
\ee
one can write the transverse-transverse interference cross section as
\be
\frac{d\sigma_{TT}}{dt}\,\simeq\,-\frac1{\kappa}\Big[|{\cal M}^N_{0+,++}|^2
                                            - |{\cal M}^U_{0+,++}|^2\Big]
\label{eq:sigTT}
\ee
where $\kappa$ is a phase-space factor. Thus, from the CLAS data one learns
that $|{\cal M}^N_{0+,++}|\gg |{\cal M}^U_{0+,++}|$ and is also large in comparison
to the amplitudes \req{eq:twist2}.  

In passing I remark that the  combinations \req{eq:N-U} are special cases of 
natural ($N$) and unnatural parity ($U$) combinations. They satisfy the 
symmetry relations
\be
{\cal M}^{N,U}_{0\nu',-\mu\nu} \,=\,\mp (-1)^\mu {\cal M}^{N,U}_{0\nu',\mu\nu}
\label{eq:NU-parity}
\ee
as a consequence of parity conservation~\footnote{
     Parity conservation leads to the relation
     ${\cal M}_{0-\nu',-\mu-\nu}=-(-1)^{\mu-\nu+\nu'} {\cal M}_{0\nu',\mu\nu}$. An analogous
     relation holds for the subprocess amplitudes.}. 
%

\section{Transversity}
\label{sec:1}
In Fig.\ \ref{fig:2} a typical Feynman graph for pion production is depicted
with helicity labels for the amplitude ${\cal M}_{0-,++}$. Angular momentum
conservation forces  both the subprocess amplitude and the nucleon-parton matrix 
element to vanish $\propto \sqrt{-t'}$ for $t'\to 0$ for any contribution to 
${\cal M}_{0-,++}$ from the usual helicity non-flip GPDs $\widetilde H$ and 
$\tilde E$. This  result is in conflict with the HERMES data on $A_{UT}^{\sin\phi_s}$. 
However, there is a second set of GPDs, the helicity-flip or transversity ones, 
$H_T, \widetilde{H}_T, E_T, {\tilde E}_T$ \ci{hoodbhoy98,diehl01} for which the 
emitted and reabsorbed partons have opposite helicities. In \ci{GK6,GK5} (see 
also \ci{liuti09}), it has been suggested that contributions from the 
transversity GPDs are responsible for the above mentioned experimental phenomena. 
Assuming handbag factorization for the $\gamma_T^*\to\pi$ amplitudes, they read 
($\mu=\pm 1$)
\ba
M_{0+\mu +} &=& e_0 \frac{\sqrt{-t^\prime}}{4m} \int^1_{-1} dx\left\{
  \big({\cal H}_{0+\mu -} - {\cal H}_{0-\mu +}\big)\big(\bar{E}_T-
                                         \xi\widetilde{E}_T\big)\right. \nn\\
        &&\left.\hspace*{0.15\tw}+ \big( {\cal H}_{0+\mu -} + {\cal H}_{0-\mu +}\big)
               \big(\widetilde{E}_T -\xi E_T\big) \right\}   \nn\\
M_{0-\mu +} &=& e_0 \sqrt{1-\xi^2} \int^1_{-1} dx \left\{{\cal H}_{0-\mu +} \Big[H_T   
            + \frac{\xi}{1-\xi^2} 
                     \big(\widetilde{E}_T -  \xi E_T\big)\Big] \right. \nn\\
            && \left. \hspace*{0.17\tw}+ \big({\cal H}_{0+\mu -} - { \cal H}_{0-\mu +}\big)
              \frac{t^\prime}{4m^2} \widetilde{H}_T \right\}\,.
\label{eq:twist3}
\ea
With the help of parity conservation one can easily convince oneself that the 
nucleon helicity non-flip amplitude consists of a natural and a unnatural parity 
part. The natural parity part is related to the GPD 
$\bar{E}_T\equiv 2\widetilde{H}_T + E_T$ with corrections~\footnote{
      The GPD ${\tilde E}_T$ is an odd function of $\xi$ as a consequence of time
      reversal invariance \ci{diehl01}.}
of order $\xi^2$ from ${\tilde E}_T$. The unnatural part is proportional to $\xi$.
The amplitude $M_{0-\mu +}$ is more complicated. There is a contribution that does 
not have a definite parity behavior. It is dominated by $H_T$ with corrections of
order $\xi^2$ from $E_T$ and $\widetilde{E}_T$. It contributes to $M_{0-++}$ 
while its contribution to $M_{0--+}$ is suppressed by $t/Q^2$ from the double-flip
subprocess amplitude $H_{0-,-+}$. In addition there is a natural parity contribution 
to ${\cal M}_{0-,\mu +}$ from $\widetilde{H}_T$ which is suppressed by 
$t^\prime/(4m^2)$. Thus, the $\gamma^*_T\to\pi$ amplitudes advocated for in 
\ci{GK6,GK5} 
\ba
{\cal M}_{0+,\mu +} &=& e_0 \frac{\sqrt{-t^\prime}}{4m} \int^1_{-1} dx 
                                             {\cal H}_{0-,++}\bar{E}_T \nn\\
{\cal M}_{0-,++} &=& e_0 \sqrt{1-\xi^2} \int^1_{-1} dx {\cal H}_{0-,++}H_T  \nn\\  
{\cal M}_{0-,-+}&=& 0
\label{eq:twist3-sim}
\ea    
are justified within the handbag approach at least at small $\xi$ and 
$-t^\prime$. There is only one subprocess amplitude common to the transverse 
amplitudes. These amplitudes meet the main features of the experimental data
discussed in Sect.\ \ref{intro}.

\section{The subprocess amplitude $H_{0-,++}$}
\label{sec:2}
As can be seen from Fig.\ \ref{fig:2}
quark and antiquark forming the pion have the same helicity. Hence, a twist-3
pion wave function is required. There are two twist-3 distribution amplitudes,
a pseudoscalar one, $\Phi_P$ and a pseudotensor one, $\Phi_\sigma$. Assuming the   
three-particle contributions to be strictly zero, one obtains the twist-3 
distribution amplitudes \ci{braun90}
\be
\Phi_P\equiv 1    \qquad \qquad \Phi_\sigma=6\tau (1-\tau)
\label{eq:PDA}
\ee
from the equation of motion. Here $\tau$ is the momentum fraction the quark in 
the pion carries with respect to the pion momentum. The subprocess amplitude 
${\cal H}_{0-,++}$ is computed to lowest-order of perturbation theory (a typical 
Feynman graph is shown in Fig.\ \ref{fig:2}). It turns out that the pseudotensor 
term is proportional to $t/Q^2$ and consequently neglected. The pseudoscalar 
term is non-zero at $t=0$ but infrared singular. In order to regularize this 
singularity the modified perturbative approach is used in \ci{GK6,GK5} in which 
quark transverse momentum, $k_\perp$, are retained in the subprocess while the 
emission and reabsorption of partons from the nucleons is assumed to happen 
collinear to the nucleon momenta \ci{GK3}. In this approach the subprocess 
amplitude reads
\ba
{\cal H}_{0-,++} &=&\frac{2}{\pi^2} \frac{C_F}{\sqrt{2N_C}}\mu_\pi 
        \int d\tau d^2{\bf k}_\perp \Psi_{\pi P}(\tau, k_\perp) \alpha_s(\mu_R) \nn\\
      &\times & \Big[\frac{e_u}{x-\xi+i\epsilon}
                     \frac1{\bar{\tau}\frac{x-\xi}{2\xi}Q^2-k_\perp^2+i\epsilon} \nn\\
       && - \frac{e_d}{x+\xi-i\epsilon}
                     \frac1{-\tau\frac{x+\xi}{2\xi}Q^2-k_\perp^2+i\epsilon}\Big]\,.
\label{eq:twist3-sub}
\ea
Here~\footnote{
     This is the subprocess amplitude for $\pi^+$ production. For the case of the
     $\pi^0$ the quark charges have to be taken out; they appear in the flavor
     combination of the GPDs.}
, $\Psi_{\pi P}$ is a light-cone wave function for the pion which is 
parametrized as~\footnote{
      It may seem appropriate to use an $l_3=\pm 1$ wave function for a particle 
      moving along the 3-direction. Such a momentum space wave function is 
      proportional to $k^{\pm}_\perp=k^1_\perp\pm ik^2_\perp$ \ci{yuan04}. However, 
      its collinear reduction leads to the pseudotensor distribution amplitude; 
      the pseudoscalar one is lacking in this wave function.} 
\be
\Psi_{\pi P}\,=\,\frac{16\pi^{3/2}}{\sqrt{2N_C}} f_\pi a_P^3|{\bf k}_\perp|
                                                    \exp{[-a_P^2k_\perp^2]}\,.
\ee
Its associated distribution amplitude is the pseudoscalar one given in \req{eq:PDA}. 
For the transverse-size parameter, $a_P$, the value $1.8\,\gev$ is adopted and
$f_\pi (=0.132\,\gev)$ denotes the pion decay constant. The parameter $\mu_\pi$ 
in \req{eq:twist3-sub} is related to the chiral condensate
\be
\mu_\pi \,=\, \frac{m_\pi^2}{m_u+m_d}
\ee
where $m_\pi$ is the pion mass while $m_u$ and $m_d$ denote current quark masses. 
In \ci{GK6,GK5} a value~\footnote{
     Taking the quark masses from \ci{PDG}, one finds $\mu_\pi=2.6^{+0.52}_{-0.15}\,\gev$
     while QCD sum rule analyses, e.g. \ci{ball98}, favor the value 
     $1.6\pm 0.2\,\gev$.}
of $2\,\gev^2$ is taken for $\mu_\pi$. The contributions from transversely
polarized photons which are of twist-3 accuracy, are parametrically suppressed by
$\mu_\pi/Q$ as compared to the $\gamma^*_L\to\pi$ amplitudes
\req{eq:twist2}. However, for values of $Q$ 
accessible in current experiments $\mu_\pi/Q$ is of order 1. 

The amplitude \req{eq:twist3-sub} is Fourier transformed to the impact parameter 
space (after partial fraction decomposition) and the integrand is 
multiplied by a Sudakov factor, $\exp{[-S]}$, which represents gluon radiation in 
next-to-leading-log approximation using resummation techniques and having recourse 
to the renormalization group \ci{li-sterman}. The Sudakov factor cuts off  
the $b$-integration at $b_0=1/\Lambda_{\rm QCD}$. In the modified perturbative
approach the renormalization and factorization scales are 
$\mu_R={\rm max}[\tau Q, (1-\tau)Q,1/b]$ and $\mu_F=1/b$, respectively. In 
\ci{GK6,GK5} the modified appoach is analogously applied to the 
$\gamma^*_L\to\pi$ amplitudes. The Sudakov factor guarantees the emergence of the  
twist-2 result for $Q^2\to\infty$.

\section{The pion pole}
\label{sec:3}
\begin{figure*}
\centering
\includegraphics[width=0.46\tw]{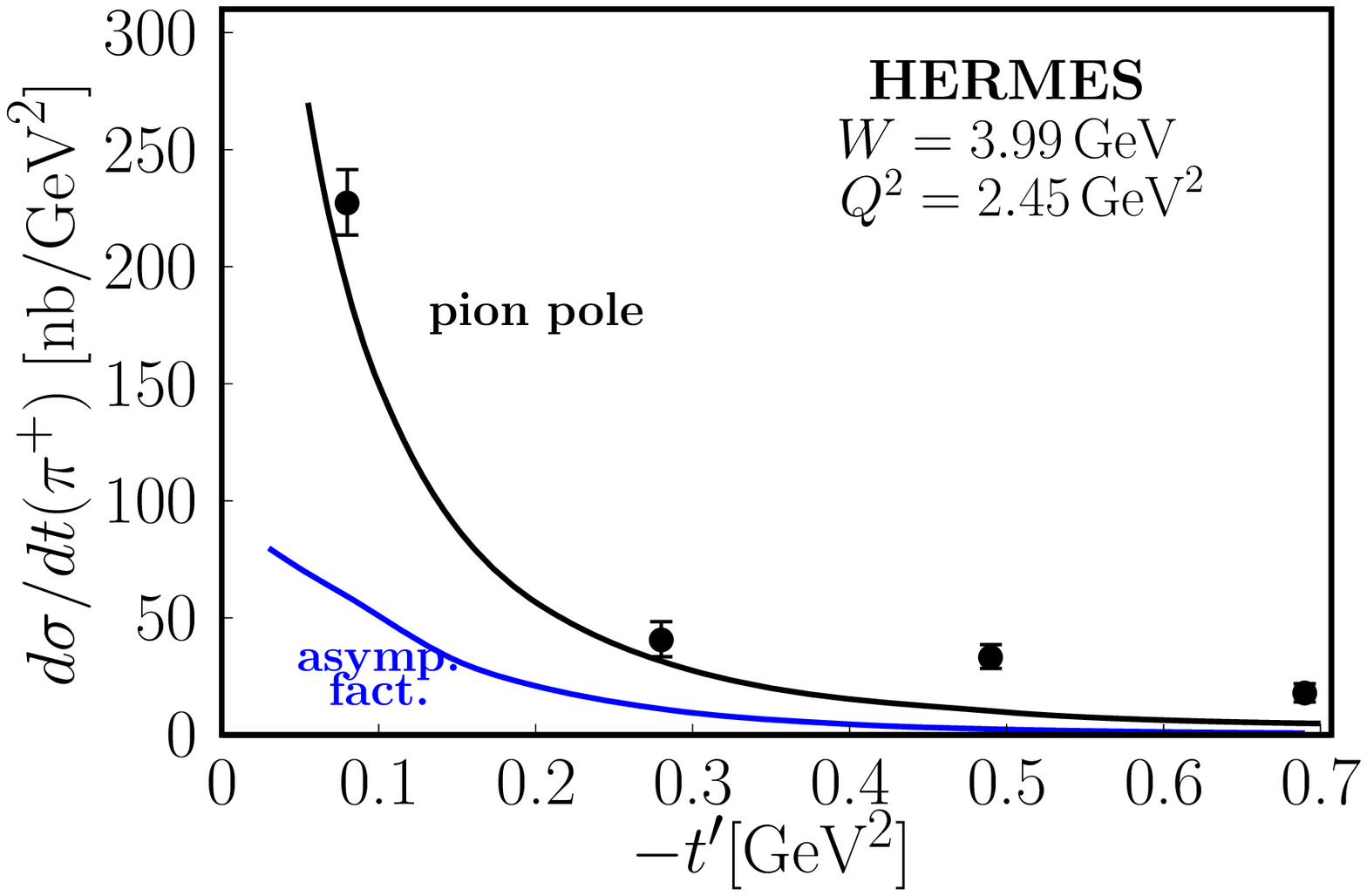}\hspace*{0.05\tw}
\includegraphics[width=0.38\tw]{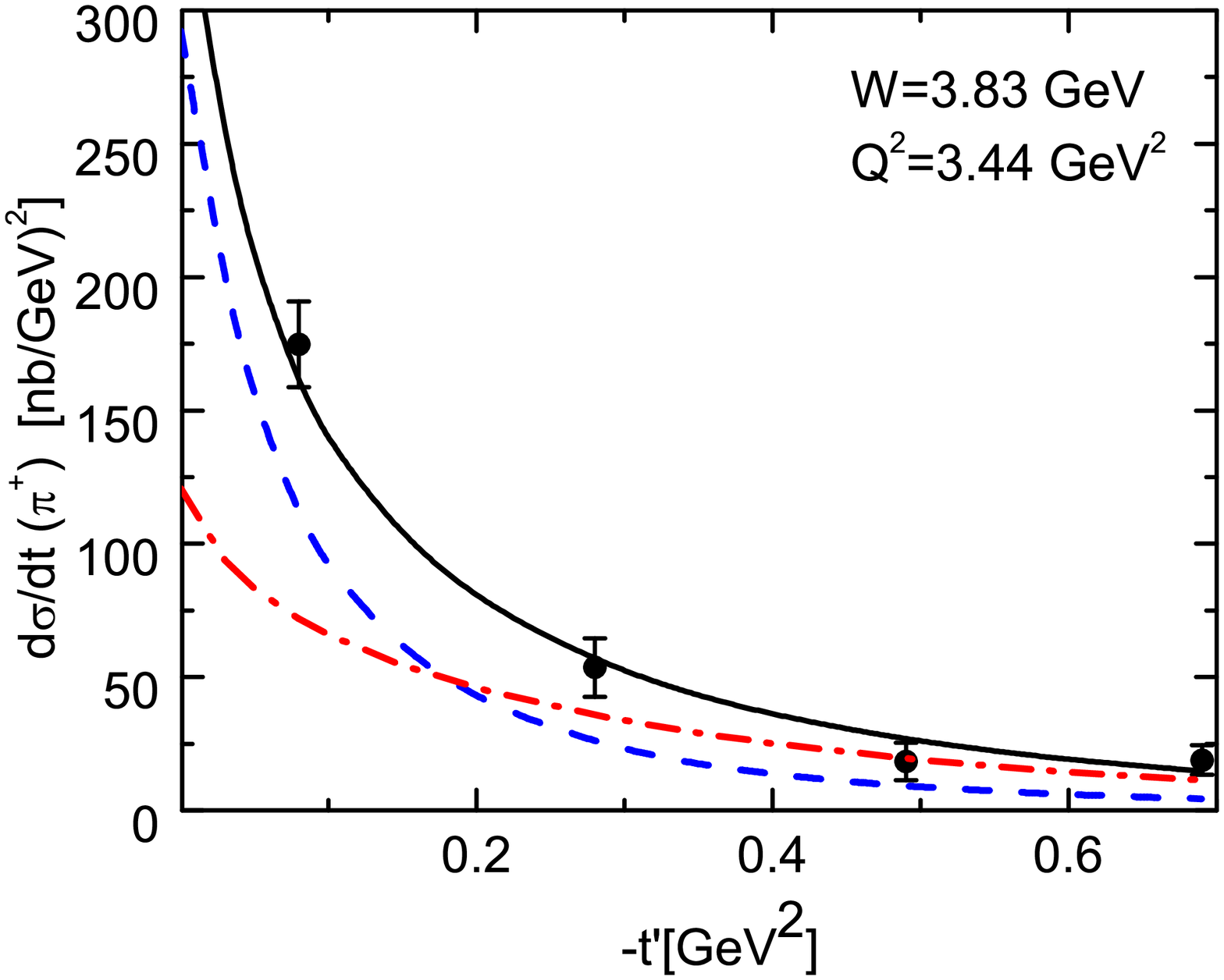}
\caption{\label{fig:3} The unseparated $\pi^+$ cross section versus $-t'$.
The lines represent the pion pole contribution \req{eq:pole-contr} with 
$F_\pi^{\rm pert}$ replaced by the experimental value and the leading-twist result. 
Data are taken from \ci{hermes08}. }
\caption{\label{fig:4} The unseparated $\pi^+$ cross section. Data taken from
\ci{hermes08}. The solid (dashed, dash-dotted) curve represents the results of 
\ci{GK5} for the unseparated (longitudinal, transverse) cross section.} 
\end{figure*}

A special feature of $\pi^+$ production is the appearance of the pion pole. As has 
been shown in \ci{mankiewicz99} the pion pole is part of the GPD $\tilde E$
\be
{\tilde E}^u_{\rm pole}\,=\,-{\tilde E}^d_{\rm pole}\,=\, \Theta(\xi -|x|)\frac{F_P(t)}{4\xi}
                         \Phi_\pi\big(\frac{x+\xi}{2\xi}\big)
\ee
where $F_P$ is the pole contribution to the pseudoscalar form factor of the nucleon
which, with the help of PCAC and the Goldberger-Treiman relation, can be written as
\be
F_P(t)\,=\,\int^1_{-1} dx\big[{\tilde E}^u-{\tilde E}^d\big] \,=\,
              -2 m \frac{f_\pi}{F_\pi(Q^2)} \frac{\varrho_\pi}{t-m_\pi^2}
\ee
with
\be
\varrho_\pi \,=\, \sqrt{2} g_{\pi NN}F_{\pi NN}(t)F_\pi(Q^2) \,.
\ee
The coupling of the exchanged pion to the nucleons is described by the coupling 
constant $g_{\pi NN} (=13.1\pm 0.2)$ and a form factor parametrized as a monopole with
a parameter $\Lambda_N (=0.44\pm 0.07$); $F_\pi$ denotes the electromagnetic form factor
of the pion. The convolution of ${\tilde E}_{\rm pole}$ with the subprocess amplitude 
${\cal H}_{0\lambda,0\lambda}$ can be worked out analytically. The result leads to the 
following pole contribution to the longitudinal cross section
\be
\frac{d\sigma_L^{\rm pole}}{dt}\,=\,4\pi\frac{\alpha_{em}}{\kappa}\frac{-t}{(t-m^2_\pi)^2}
                                    Q^2\varrho^2_\pi\,.  
\label{eq:pole-contr}
\ee
However, in this calculation the pion form factor is only the leading-order
perturbative contribution to it which is known to underestimate the experimental 
form factor \ci{F-pi-08} substantially, and consequently the $\pi^+$ cross section, see 
Fig.\ \ref{fig:3}. In \ci{GK6,GK5} the perturbative result for $F_\pi$ was therefore    
replaced by its experimental value. This prescription is equivalent to computing the 
pion pole contribution as a one-particle exchange~\footnote{
     For $Q^2\gg -t$ the virtuality of the exchanged pion can be neglected. The 
     $\gamma^*\to \pi\pi^*$ vertex is therefore the pion form factor of the pion.}. 
With this replacement one sees that the pole term controls the $\pi^+$ cross section 
at small $-t'$, see Fig.\ \ref{fig:3}. A detailed discussion of the pion pole 
contribution can be found in \ci{FGHK15}. It also plays a striking role in 
electroproduction of $\omega$ mesons \ci{GK9}.  
%

\section{Phenomenology}
\label{sec:4}
In order to make predictions and comparison with experiment the GPDs are needed.
In \ci{GK6,GK5,GK3} the GPDs are constructed with the help of the 
double-distribution representation. According to \ci{musatov00} a double 
distribution is parametrized as a product of a zero-skewness GPD and an 
appropriate weight function that generates the skewness dependence. The 
zero-skewness GPD itself is composed of its forward limit, $K(x,\xi=t=0)=k(x)$, 
multiplied by an exponential in $t$ with a profile function, $f(x)$, parametrized 
in a Regge-like manner
\be
f(x)\,=\,-\alpha'\ln{x} + B
\ee
at small $-t$. The GPD $\widetilde H$ at $\xi=0$ is taken from a recent analysis
of the nucleon form factors \ci{DK13} while $\tilde E$ is neglected. The forward 
limit of the transversity GPD $H_T$ is given by the transversity parton 
distributions, known from an analysis of the azimuthal asymmetry in semi-inclusive 
deep inelastic lepton-nucleon scattering and inclusive two-hadron production in 
electron-positron annihilations \ci{anselmino08}.It turns out, however, that this 
parametrization underestimates $H_T$. Its moments are about $40\%$ smaller than 
lattice QCD results \ci{gockeler05} and it leads to a very deep dip in the forward 
$\pi^0$ cross section which is at variance with experiment \ci{clas-pi0}. Therefore, 
the normalization of $H_T$ is adjusted to the lattice QCD moments but the 
$x$-dependence of the transversity distributions is left unchanged. The forward 
limit of $\bar{E}_T$ is parametrized like the usual parton densities
\be
\bar{e}_T(x)\,=\,N x^{-\alpha}(1-x)^\beta
\ee
with $\alpha=0.3$ for both $u$ and $d$ quarks and $\beta^u=4$, $\beta^d=5$.
The normalization as well as the parameters $\alpha'$ and $B$ for each of the 
transversity GPDs are estimated by fits to the HERMES data on the $\pi^+$ cross 
section \ci{hermes08} and to the lattice QCD results \ci{gockeler07} on the 
moments of $\bar{E}_T$.

An example of the results for the $\pi^+$ cross section is shown in Fig.\ 
\ref{fig:4}, typical results for $\pi^0$ production in Fig.\ \ref{fig:1}. At small 
$-t'$ the $\pi^+$ cross section is under control of the pion pole as discussed in 
Sect.\ \ref{sec:3}. The contribution from $\widetilde H$ to the longitudinal 
cross section is minor. The transverse cross section, although suppressed by 
$\mu_\pi^2/Q^2$, is rather large and even dominates for $-t'\gsim 0.2\,\gev^2$.
It is governed by $H_T$, the contribution from $\bar{E}_T$ is very small. This
fact can easily be understood considering the relative sign of $u$ and $d$
quark GPDs. For $H_T$ they have opposite signs but the same sign for $\bar{E}_T$.
Moreover, $\bar{E}^u_T$ and  $\bar{E}^d_T$ are of similar size~\footnote{
      In \ci{burkardt06} it has been speculated that $\bar{E}_T$ is linearly
      related to the Boer-Mulders function. Indeed both the functions show the 
      same pattern \ci{barone09}. This pattern is also supported by large-$N_c$
      considerations \ci{weiss16}.}. 
Since the GPDs contribute to $\pi^+$ production in the flavor combination
\be
K^{\pi^+} \,=\, K^u - K^d
\ee
it is obvious that the contributions from $\bar{E}_T^u$ and $\bar{E}^d$ cancel 
each other to a large extent in contrast to those from $H_T$.    

\begin{figure*}
\centering
\includegraphics[width=0.24\tw]{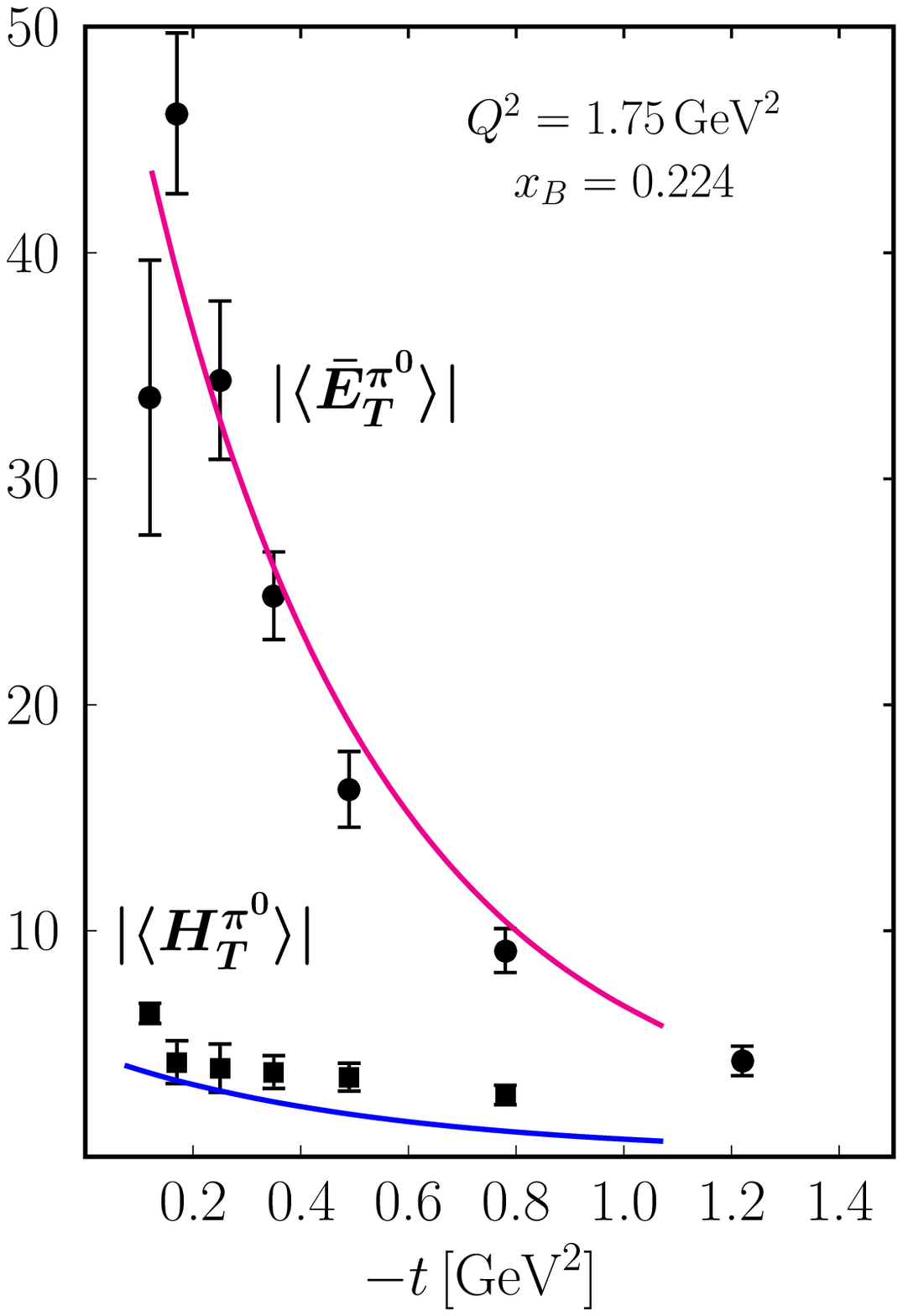}
\hspace*{0.07\tw}
\includegraphics[width=0.24\tw]{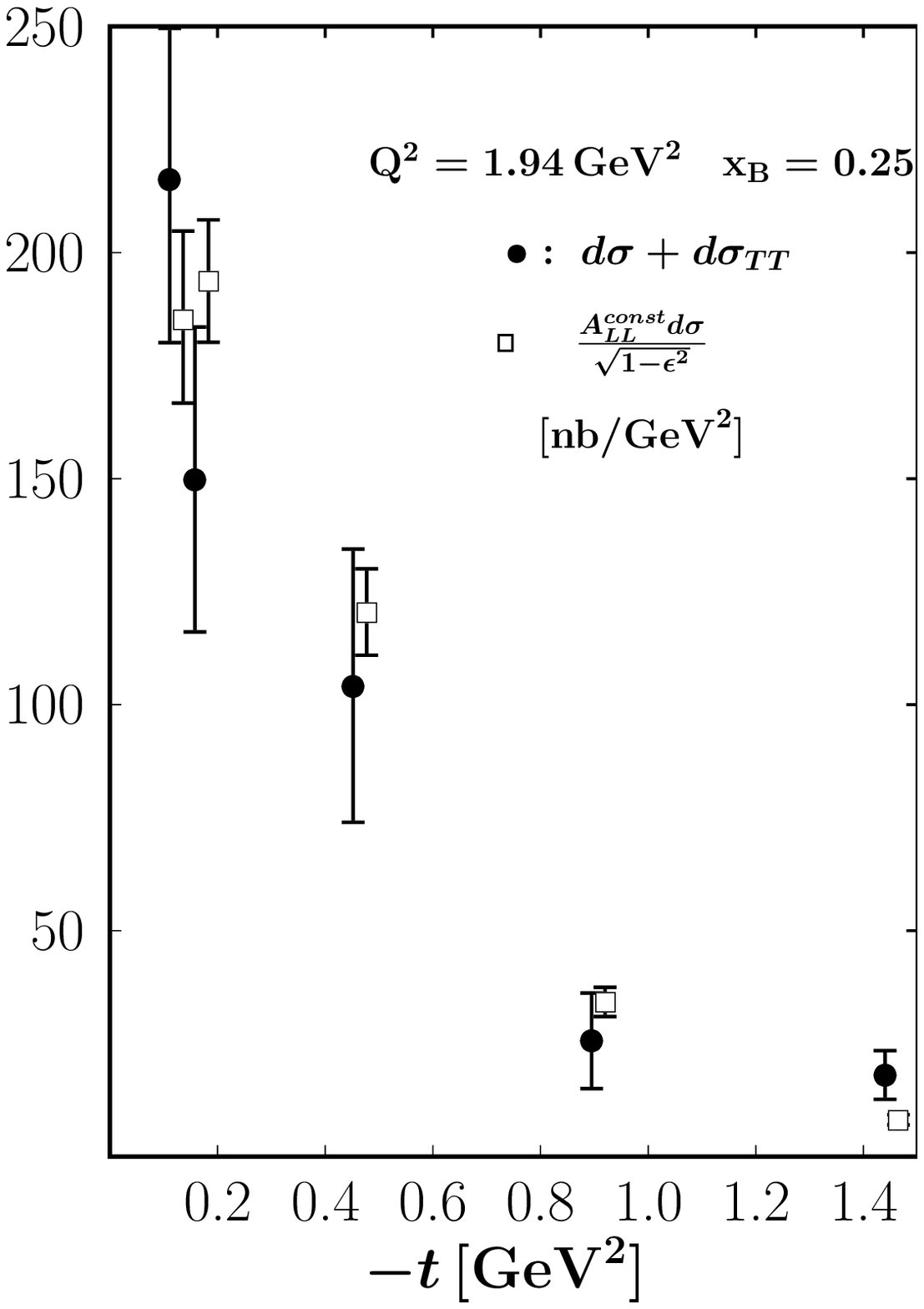}
\caption{\label{fig:5} The convolutions of $\bar{E}_T$ and $H_T$ extracted from the
CLAS data on $\pi^0$ leptoproduction \ci{clas-pi0} and the estimates from \ci{GK6,GK5}.}
\caption{\label{fig:6} Testing the relation \req{eq:ALL-relation}. Data taken from
\ci{clas-pi0} and \ci{kim15}.}
\end{figure*}

For $\pi^0$ production the situation is reversed since the GPDs appear now
in the flavor combination
\be
K^{\pi^0}\,=\,\frac1{\sqrt{2}} \big(e_u K^u - e_d K^d\big)\,.
\label{eq:pi0-flavor}
\ee 
Therefore, $\bar{E}^u_T$ and $\bar{E}^d_T$ add while there is a partial cancellation
between $H^u_T$ and $H^d_T$. As can be seen from Eqs.\ \req{eq:NU-parity} and 
\req{eq:twist3-sim} the $\bar{E}_T$-contribution is of the natural-parity type
and, hence, makes up the transverse-transverse interference cross section 
\req{eq:sigTT}, see Fig.\ \ref{fig:1}. The transverse cross section, $d\sigma_T/dt$ 
receives contribution from both $H_T$ and  $\bar{E}_T$. However, the sum
\be
\frac{d\sigma_T}{dt} + \frac{d\sigma_{TT}}{dt}\,\simeq\,\frac1{2\kappa}|{\cal M}_{0-,++}|^2 
\ee
is only fed by $H_T$. According to \ci{GK6,GK5} 
\be
\frac{d\sigma_L}{dt} \ll  \frac{d\sigma_T}{dt}\,.
\label{eq:LT}
\ee
Hence, to a good approximation the transverse cross section equals the unseparated 
one. The prediction \req{eq:LT} is consistent with the very small 
longitudinal-transverse interference cross section, see Fig.\ \ref{fig:1}. It is to 
be emphasized that this prediction is what is expected in the limit $Q^2\to 0$ and 
not for $Q^2\to \infty$. With the approximation $d\sigma_T\simeq d\sigma$ one can 
directly extract the convolutions of $H_T^{\pi^0}$ and $\bar{E}_T^{\pi^0}$ from the data 
on the $\pi^0$ cross sections. Results for the convolutions are displayed in Fig.\ 
\ref{fig:5} at $Q^2=1.75\,\gev^2$ and $x_B=0.224$ and compared to the estimates
made in \ci{GK6,GK5}. There is a further test of the set of amplitudes 
\req{eq:twist3-sim}: The 'constant' modulation of the asymmetry $A_{LL}$ measured 
with longitudinally polarized beam and target by the CLAS collaboration \ci{kim15} 
for $\pi^0$ production, is related to the cross sections by
\be
\frac{A_{LL}^{\rm const}}{\sqrt{1-\varepsilon}}\frac{d\sigma}{dt} \simeq 
                \frac{d\sigma_T}{dt} + \frac{d\sigma_{TT}}{dt}
         \simeq \frac{d\sigma}{dt} + \frac{d\sigma_{TT}}{dt}
\label{eq:ALL-relation}
\ee
($\varepsilon$ denotes the ratio of the longitudinal and transversal photon fluxes). 
A violation of this relation would indicate contributions from other transversity 
GPDs, especially from 
${\widetilde H}_T$ (see \req{eq:twist3}). In Fig.\ \ref{fig:6} the right and left
hand sides of \req{eq:ALL-relation} are compared to each other. Within admittedly
large errors there is agreement except, perhaps, at the largest values of $-t$. Of
course small contributions from other transversity GPDs cannot be excluded.

Data on $\pi^0$ production off neutrons as is planned to measure by the 
Jefferson Laboratory Hall A collaboration, will improve the flavor separation
of the GPDs since they now appear in the combination
\be
K^{\pi^0}_{\rm neutron}\,=\,\frac1{\sqrt{2}} \big(e_u K^d - e_d K^u\big)\,.       
\ee
In analogy to the proton case (see Fig. \ref{fig:6}) one may extract the 
convolutions of $H_T$ and $\bar{E}_T$ from future data. According to the present 
estimate $\bar{E}_T$ for $u$ and $d$ quarks have about the same size. If this is 
correct the neutron/proton ratio of $d\sigma_{TT}$ is about one. On the other hand, 
for $d\sigma_T+d\sigma_{TT}$ the neutron/proton ratio is expected to be much smaller 
than one because of the properties of $H_T$.  

\begin{figure*}
\centering
\includegraphics[width=0.4\tw]{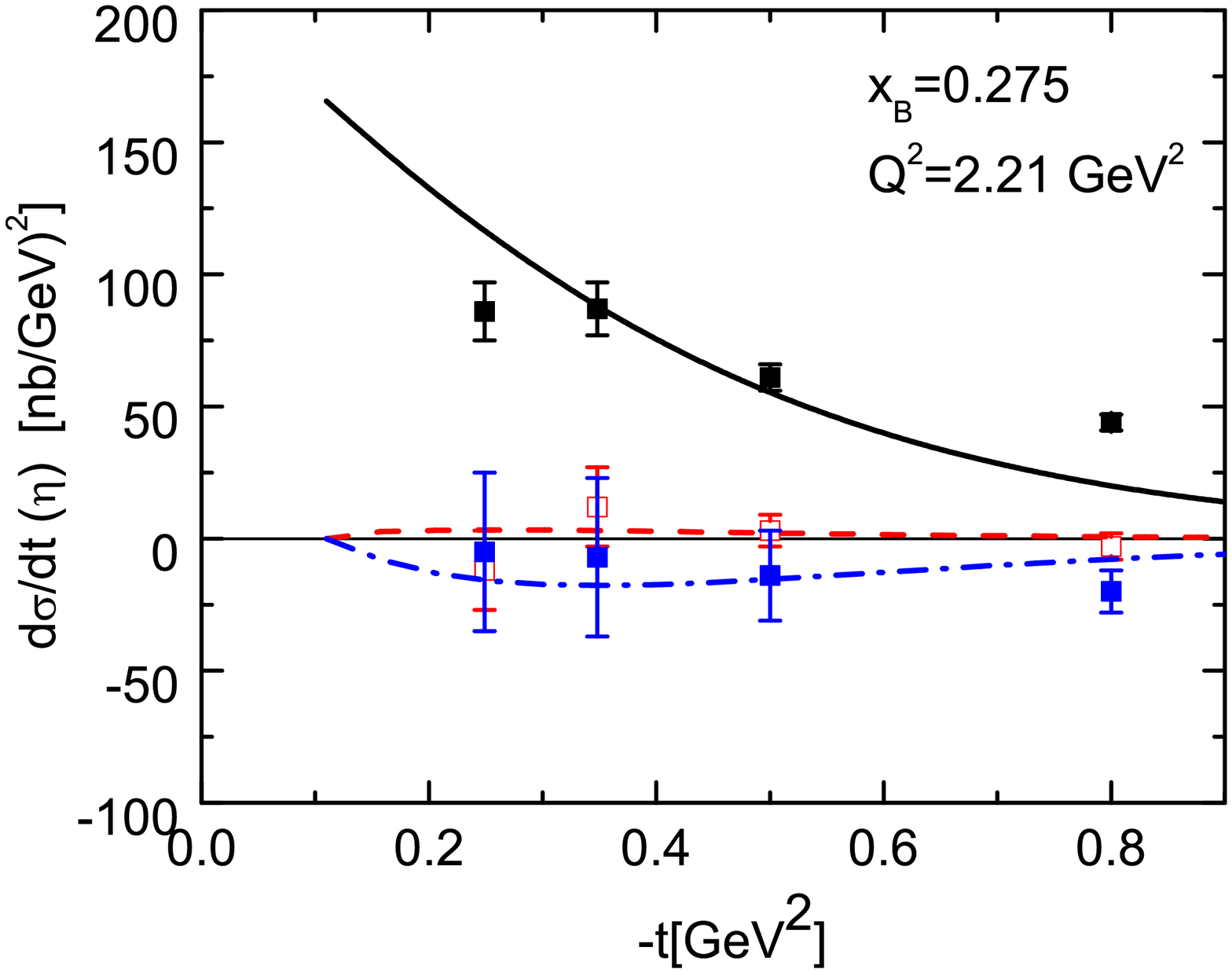}
\hspace*{0.05\tw}
\includegraphics[width=0.37\tw]{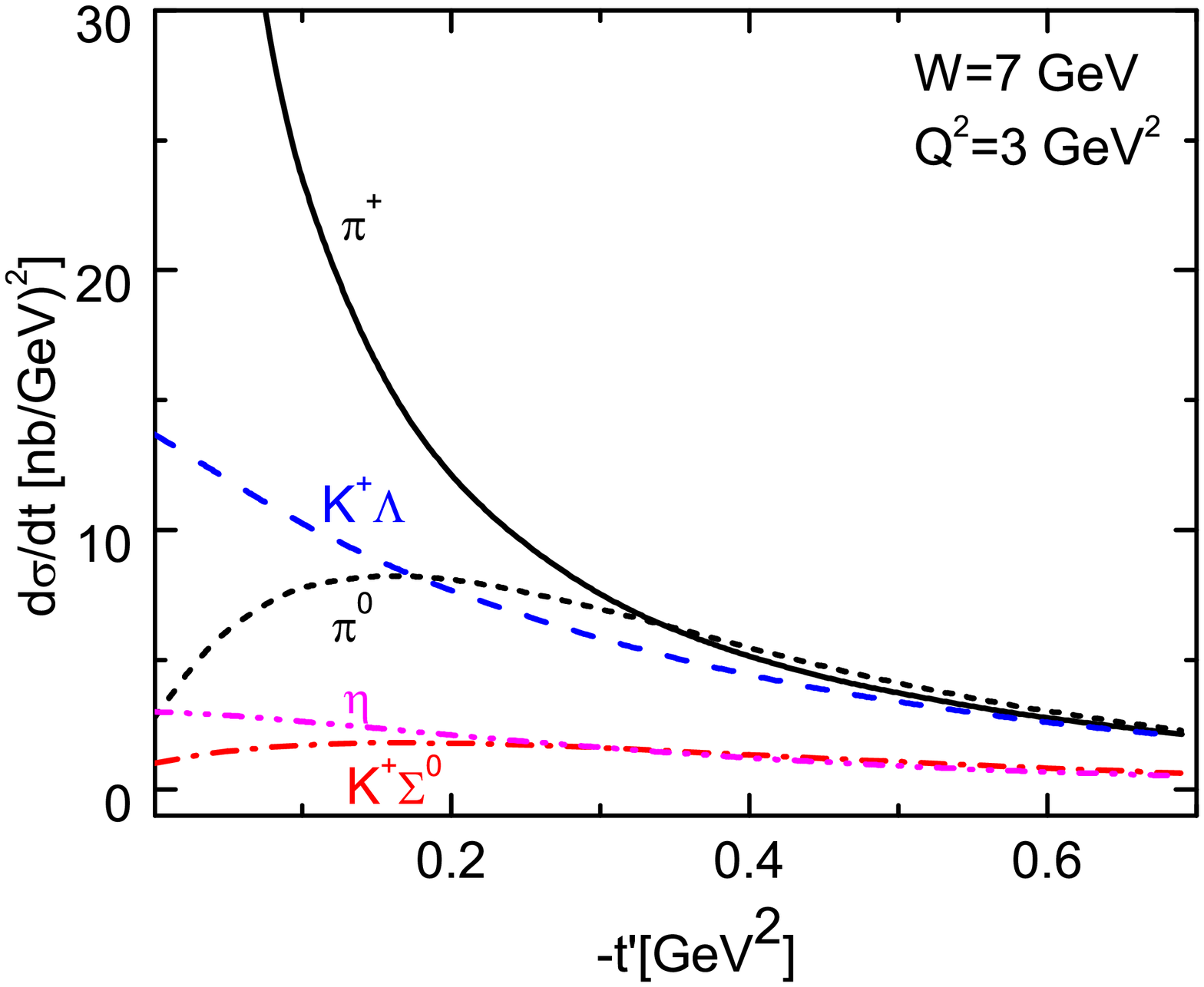}
\caption{\label{fig:7} The $\eta$ production cross sections. Predictions are from
\ci{GK6}, preliminary data are from \ci{kubarowsky10}.} 
\caption{\label{fig:8} Predictions for the unseparated cross sections of various 
pseudoscalar meson channels \ci{GK6}.} 
\end{figure*}

The transversity GPDs play a similarly prominent role in leptoproduction of other
pseudoscalar mesons. In Fig.\ \ref{fig:7} predictions for the $\eta$ cross sections 
are shown. Except in the proximity of the forward direction the unseparated cross
section for $\eta$ production is considerably smaller than the $\pi^0$ one. In fact 
ratio $d\sigma(\eta)/d\sigma(\pi^0)$ amounts to about a third in fair agreement 
with the preliminary CLAS data on $\eta$ production \ci{kubarowsky10}. In the small
$-t'$ region where the GPD $H_T$ takes the lead, the $\eta/\pi^0$ ratio amounts to
about 1 \ci{GK6}. The behavior of the ratio found in \ci{GK6} is in sharp 
contrast with earlier speculations \ci{eides} that the ratio is larger than 1 for all 
$t'$.
One can easily understand the  results for the $\eta/\pi^0$ ratio by considering 
again the relative signs and sizes of the $u$ and $d$-quark GPDs. Under the plausible
assumption $K^s=K^{\bar{s}}$ only the $u$ and $d$-quark GPDs in the combination~\footnote{
       Due to this assumption the flavor singlet and octet combinations are related by
       $K^{(1)}=\sqrt{2}K^{(8)}$.}
\be
K^\eta \simeq \frac1{\sqrt{6}}\big(e_u K^u + e_d K^d\big)\,.
\label{eq:eta-flavor}
\ee
contribute to $\eta$ production.  
With regard to the different signs in \req{eq:eta-flavor} and \req{eq:pi0-flavor} 
it is evident that $H_T$ plays a more important role for $\eta$ than for $\pi^0$ 
production while for $\bar{E}_T$ the situation is reversed with the consequence of 
a large $\eta/\pi^0$ ratio for $t'\to 0$ and a small ratio otherwise. In the 
evaluation of the $\eta$ cross section $\eta - \eta'$ mixing is to be taken into 
account \ci{FKS1}. The $\eta$ cross sections are shown in Fig.\ \ref{fig:7}. The 
transverse-transverse cross section is now very small because of the strong 
cancellation between $\bar{E}_T^u$ and $\bar{E}_T^d$.

The handbag approach can straightforwardly be generalized to the production of
Kaons \ci{GK6}. Some results on the cross sections for various pseudoscalar meson
channels are shown in Fig.\ \ref{fig:8}. 

%

\section{Summary}
\label{summary}
In this article the present status of the analysis of hard exclusive 
leptoproduction of pions and other pseudoscalar mesons within the handbag 
approach is reviewed. The present GPD parametrizations are to be considered
as estimates which however reproduces the main features of the data.
A detailed fit to all pion data is pending.
The surprising result is the dominance of $\gamma^{\,*}_T\to \pi$ transitions. 
The leading-twist contribution is small, in particular for $\pi^0$ production.
The ultimate justification of this observation would be a measurement of the
unseparated cross sections. The JLab Hall A collaboration has done this for
$\pi^0$ production. The experiment is under analysis. 

The statement which has been mentioned in many papers and talks, that 
from pion leptoproduction we learn about the GPDs $\widetilde H$ and 
$\tilde E$ which was state of the art 10-15 years ago, is to be revised:
from pion leptoproduction we learn about contributions from transversely
polarized photons and in particular about the transversity GPDs $H_T$ and
$\bar{E}_T$.



\end{document}